\documentstyle[seceq,supplement,epsf]{ptptex}
 
\title{Chaos in the Thermodynamic Limit}

\author{
Vito {\sc Latora}$^{1}$\footnote{ E-mail: latora@ct.infn.it},
Andrea {\sc Rapisarda}$^{1}$\footnote{ E-mail: rapisarda@ct.infn.it}
and Stefano {\sc Ruffo}$^{2}$\footnote{ E-mail: ruffo@avanzi.de.unifi.it}
}
 
\inst{
$^1$ Dipartimento di Fisica, Universit\'a di Catania, 
Corso Italia 57,
\\

and  INFN sezione di Catania,  I-95129 Catania ITALY
\\ 
$^2$ Dipartimento di Energetica ``S. Stecco'',
Universit\'a di Firenze, via S. Marta 3, 
\\ INFM and INFN sezione di Firenze I-50139 Firenze, Italy}

\recdate{
}
 
\abst{We study chaos in the Hamiltonian Mean Field model (HMF), 
a system with many degrees of freedom in which $N$ classical 
rotators are fully coupled. 
We review the most important results on the dynamics and 
the thermodynamics of the HMF, and in particular we focus on 
the chaotic properties.
We study the Lyapunov exponents and the 
Kolmogorov--Sinai entropy, namely their dependence on the number
of degrees of freedom and on energy density, both for the 
ferromagnetic and the antiferromagnetic case.}

\begin{document}
 
\maketitle
 
\section{Introduction}
In systems with a few degrees of freedom the Largest Lyapunov
Exponent (LLE), which quantifies chaotic motion, is often studied
as a function of the control parameter. In many-degrees-of-freedom
systems, this can also be done and, moreover, the control
parameter may acquire a more transparent physical meaning, making
reference to a thermodynamic quantity.
There have been indeed several studies\cite{rf:pet,rf:dauxois} 
of the dependence of the LLE on energy density 
in Hamiltonian systems with short-range interactions 
(e.g. FPU lattices), which, up to now
confirm the conjecture~\cite{rf:ruelle,rf:lpr} that the LLE
reaches a finite, energy dependent, value in the thermodynamic
(large-volume) limit. Moreover, a scaling limit exists for the
full Lyapunov spectrum, which implies that the Kolmogorov-Sinai
entropy $S_{KS}$ scales with the volume.
 
Both the question of the {\it existence of a well defined 
thermodynamic limit} of LLE and $S_{KS}$, and their  
{\it dependence on energy} (or other control parameters) are open
for systems with long-range interactions.
In this paper we review the most recent results on this subject for   
a Hamiltonian model with many degrees of freedom, named 
Hamiltonian Mean Field (HMF), which describes a 
{\it fully-coupled} system of $N$ classical spins (rotators) in the 
attractive (ferromagnetic) and repulsive (antiferromagnetic) 
cases~\cite{rf:def,rf:antoni}. 
HMF has been recently thoroughly investigated both from a 
theoretical and a numerical point of view 
\cite{rf:yama,rf:prl1,rf:pd,rf:prl2},
revealing a deep link between dynamics and thermodynamics. 
Here, we discuss this relation by studying the Lyapunov exponents 
and the Kolmogorov--Sinai entropy, namely their dependence on
$N$ and on energy, both for the ferromagnetic and the 
antiferromagnetic case.

\section{The model}
 
The model describes a system of $N$ classical spins (rotators)  
${\bf m}_i=(cos \theta_i, sin \theta_i)$. 
Each  spin $i$ is characterized by the angle $0 \le \theta_i < 2\pi$ 
and the conjugate momentum $p_i$, and is {\it fully coupled} 
to all the others.   
The Hamiltonian is:  
\begin{equation}
        H(\theta,p)=K+V ~,
\end{equation}
where 
\begin{equation}
       K= \sum_{i=1}^N  \frac{{p_i}^2}{2} ~~~~~ ,
       V= \frac{\epsilon}{2N} \sum_{i,j=1}^N  [1-cos(\theta_i -\theta_j)]
~~
\end{equation}
are the kinetic and potential energy. 
The potential energy $V$ corresponds to the interaction of the 
$X-Y$ model in {\it the infinite--range mean field case}, 
and this is the reason why the model has been named 
Hamiltonian Mean Field (HMF). 
The case $\epsilon=1$ describes a ferromagnetic behavior, while 
$\epsilon=-1$ corresponds to an antiferromagnetic interaction. 
The model has a possible alternative interpretation. 
It can in fact be seen as a system of particles moving 
on a circle, the position of each particle being given 
by the angle $\theta_i$ and its momentum by $p_i$. 

The success of the Hamiltonian mean field model is based 
on the fact that both its statistical mechanics 
and its dynamics can be treated in relatively simple way. 
In fact the {\it thermodynamics} of the HMF can be derived exactly 
for $N \to \infty$ in the {\it canonical ensemble}, both 
for the ferromagnetic and for the antiferromagnetic case\cite{rf:antoni}. 
A total magnetization vector can be defined as 
${\bf M}={\frac{1}{N}}\sum_{i=1}^N {\bf m}_i$. 
The ferromagnetic system has a second--order phase transition 
from a clustered phase with $M=|{\bf M}| \neq 0$ to a disordered phase with 
$M=0$ as a function of energy or temperature. 
In the antiferromagnetic case spins tend to be opposite to each
other (interaction among the particles is repulsive) and therefore 
$M=0$ (disordered state) for any value of the temperature. 
 
On the other side the {\it dynamics} of the system can be 
investigated for a relatively large value of 
$N$ (we have considered N up to a value of 20000) 
by solving the $2N$ coupled equations of motion:  
\begin{equation} 
\dot{\theta_i}={p_i}, ~~~\dot{p_i}  = - \epsilon M sin(\theta_i - \phi ) ~~~,~~~ 
i=1,...,N~~~,
\label{eqmoto} 
\end{equation}
where $(M,\phi)$ are respectively the modulus and the phase 
of the total magnetization vector $\bf{M}$.
Each spin moves in a mean field which is in turn generated 
self--consistently by the all the other spins.
In the $N \to \infty$ limit the dynamics of the HMF can be seen 
as the interaction of a single spin with a mean field,  
and the equations are formally equivalent to those of
a perturbed pendulum. 
Solving equations (\ref{eqmoto}) corresponds to treating the 
system in the {\it microcanonical ensemble}, because the 
total energy is conserved along each dynamical trajectory
(also total momentum is a conserved quantity and is typically
fixed at zero). 
 
Hence, HMF has a very remarkable property: it is possible 
to compare the results of the canonical and microcanonical ensemble. 
Moreover the HMF is a high dimensional system  
with long--range forces where one can
explore deviations from standard 
thermodynamics~\cite{rf:celia1,rf:celia2,rf:tsallis1,rf:tsallis2,rf:tsallis}.  
HMF exhibits a very rich non--equilibrium dynamics, 
and by using the dynamical approach we have studied in detail 
the problem of the relaxation to the canonical equilibrium. 
The following is a brief review of the main results.  
 
{\it Ensemble inequivalence}:  
because of the long--range nature of the interaction, the microcanonical ensemble 
gives different predictions from the canonical ensemble\cite{rf:her71}. 
This is true both in the ferromagnetic case, where we find the presence 
of quasistationary states different from the canonical 
equilibrium~\cite{rf:prl1,rf:pd,rf:prl2},  
and in the antiferromagnetic case, where a very particular 
collective phenomenon (bi--cluster formation) appears at low energies in   
disagreement with the canonical predictions\cite{rf:def,rf:antoni}.

{\it Metastability}:
in the ferromagnetic case  microcanonical simulations 
show the presence of quasi-stationary metastable states with negative specific 
heat\cite{rf:prl1,rf:pd,rf:prl2}. 
In fact, if the system is started in far-off-equilibrium initial conditions 
(for example in a ``water bag'', i.e. putting all the rotators 
at $\theta_i=0$ and giving them a uniform distribution of 
velocities with a finite width centered around zero), 
it does not relax directly to the canonical equilibrium. Instead 
we observe a stabilization into metastable states.  
The temperature of these states are different from the canonical
one and the velocity distributions are not Gaussian\cite{rf:tsallis}. 
The metastable states are called quasistationary states because 
they have a lifetime which increases linearly with the number of 
particles $N$. They are expected to become real equilibrium 
solutions in the thermodynamic limit\cite{rf:ahr}.  

{\it Collective phenomena}: 
in the antiferromagnetic case, a bi--cluster of rotators 
at a distance $\pi$ in angle is present at very low energy 
in the microcanonical numerical simulations\cite{rf:antoni}. 
In fact, below a threshold energy, particles groups spontaneously  into two 
big clusters and oscillate maintaning   the total magnetization equal to zero.
This is a pure microcanonical result, stable also for very large
$N$ and not in agreement with the canonical ensemble \cite{rf:ruffonew},
where a disordered state with all the spins randomly oriented is predicted.  
This collective phenomenon modifies the energy--temperature 
relation at very small energies and is an effect of the 
long--range interaction\cite{rf:ruffonew}. 
 
{\it Anomalous diffusion}: 
diffusion and transport of a particle in a medium or in a fluid 
flow are characterized by the average square displacement $\sigma^2(t)$. 
In general one has 
\begin{equation}
\label{anoma}
    \sigma^2(t) \sim  t^{\alpha}
\end{equation}
with $\alpha=1$ for normal diffusion.   
All the processes with $\alpha \ne 1$ are termed anomalous 
diffusion.\cite{rf:ad1,rf:ad2,rf:ad3,rf:ad4,rf:ad5,rf:ad6} 
In our model the variance of the spin angle $\theta$ can be defined 
according to the expression:  
\begin{equation}
    \sigma_{\theta}^2(t) = < (\theta - <\theta> )^2 >
\end{equation}
where  $< ~.~>$ stands for an average over the N spins.   
Superdiffusion with $\alpha = 1.38 \pm 0.05$ is observed in the ferromagnetic 
case in the energy range $0.5<U<0.75$, i.e. slightly below the 
critical energy $U_c=0.75$\cite{rf:prl2}.
Superdiffusion is due to the presence of L\'evy flights\cite{rf:ad1},   
and after a transient regime a change to the slope $\alpha = 1$ (normal 
diffusion) is observed\cite{rf:grigo}. Normal diffusion occurs at a 
crossover time which we have found to coincide with the relaxation time to
canonical equilibrium \cite{rf:prl2}. However in the continuum limit 
diffusion is always anomalous since the crossover time and the relaxation
one diverge with $N$. These investigation confirms pioneering studies by
Kaneko and Konishi for coupled maps \cite{rf:kan,rf:kon}, and are on the same line
of investigation of refs.\cite{rf:grigo,rf:torc}
where the effect of noise or fluctuations on the diffusion is studied.
 
{\it Generalization of HMF}: 
generalizations of the model have already appeared 
in the literature. 
In the model introduced by Anteneodo and Tsallis the rotators
have been attached to the sites of a 1D lattice and 
the interaction of the HMF has been modulated by a term depending 
on the lattice distance between two spins, going like a decaying 
power-law $r^{-\alpha}$~\cite{rf:celia1,rf:celia2}.  
The 2D case of the HMF has been considered by Antoni and Torcini~\cite{rf:torc}.
Further generalizations, like 1D models with spatial modulations or 
1D models with mixed (attractive-repulsive) interactions, 
are currently under investigation. 
 
In this paper we focus mainly on the study of the chaotic dynamics in the HMF. 
Our investigations are relevant for the foundation of 
statistical mechanics and also for the study of phase transitions 
in finite size systems like for example: nuclear 
multifragmentation~\cite{rf:blr,rf:gross}, 
atomic clusters~\cite{rf:cluster1,rf:cluster2} 
and  astrophysics~\cite{rf:astro}. 
HMF, with the possibility of considering both the ferromagnetic 
and the antiferromagnetic cases, offers two different scenarios. 
In the following sections we will show that the chaotic 
properties of the system are different in these two cases.

\section{Thermodynamics: the canonical solution}
 
The HMF presents the noticeable advantage of possessing an exact solution 
in the canonical ensemble. 
Therefore  microscopic dynamics, and in particular the  
chaotic properties, can be studied in connection with the thermodynamic  
macroscopic behavior. 
 
\begin{figure}
\epsfxsize = 8 cm 
\centerline{\epsfbox{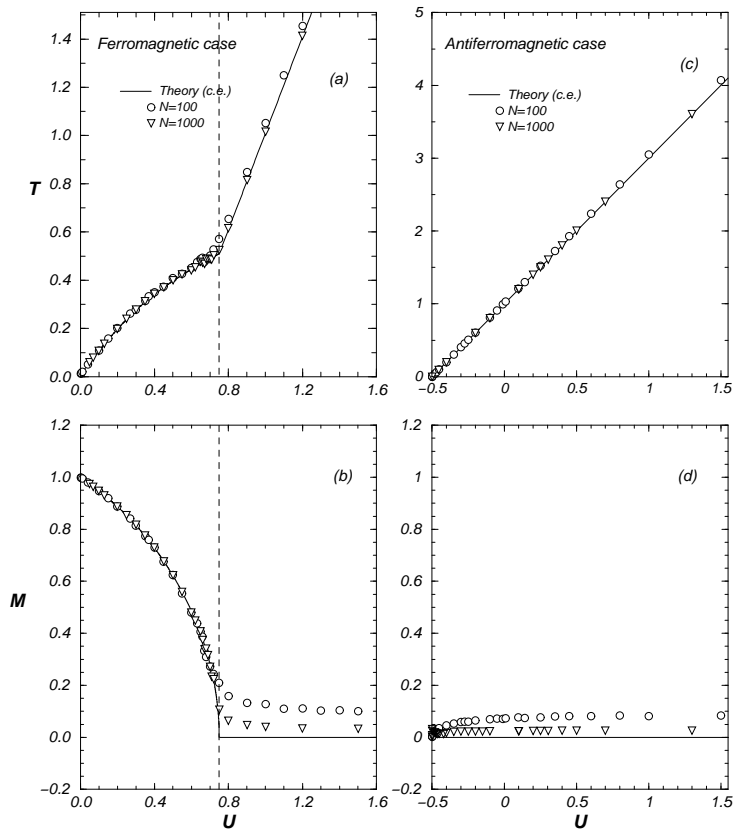}}
\caption{We plot the behavior of temperature 
and magnetization as a function of the energy per particle $U$ 
in the ferromagnetic (a)(b) and in the antiferromagnetic 
case (c)(d). Points are microcanonical numerical simulations for $N=100,1000$ while
the solid curves are the exact canonical predictions, see text.
The dashed line indicates the critical point for the ferromagnetic case.
}
\label{mari}
\end{figure}
 
In this section we discuss the canonical solution for both 
$\epsilon=+1$ (ferromagnet) and $\epsilon=-1$ (antiferromagnet).

In the ferromagnetic model the potential is attractive and the 
ground state of the system is reached at $U=0$ where all spins are 
parallel (all the rotators have the same position on the circle). 
In the antiferromagnetic model the potential is repulsive and 
the ground state, reached at $U=-1/2$,
consists in a randomly uniform distribution of the spins 
orientations. 

In the high temperature region, both in the ferromagnetic and in 
the antiferromagnetic model, the rotators are randomly distributed 
on the circle; each rotator moves uniformly around 
the circle and the modulus of $|\bf{M}|$ is equal to zero. 
Thermodynamically, when the potential is attractive the HMF 
has a second--order phase transition with order parameter $M$, 
while in the repulsive case the free energy is smooth and $M=0$ 
for any value of $T$.
The exact solution of the model in the canonical ensemble predicts 
a caloric curve given by  
\begin{equation}
U = \frac{T}{2} + \frac{\epsilon}{2}(1-M^2)~.
\label{can}
\end{equation}
The ferromagnetic case has a critical temperature $T_c=0.5$ ($U_c=0.75$), 
while in the antiferromagnetic case there is no phase transition and 
$M=0$ in equation (\ref{can}).  
In fig.1 we plot the behavior of magnetization and 
temperature as a function of the energy per particle.  
The solid curves are the canonical predictions (see eq.(\ref{can})) while the 
points are the results of the microcanonical numerical simulations 
for a system with N=100 and N=1000 (see next section).
The vertical dashed line indicates the critical point for the 
ferromagnetic case.
The figure shows that there are small deviations due to the finite size
of the system. These deviations decrease as $N^{-1/2}$. In the repulsive
case there are also deviations from the canonical equilibrium at very small
energies due to the formation of the bi-cluster (for a detailed 
discussion see ref.~\cite{rf:ruffonew})

\section{Dynamics: the microcanonical simulations}
 
We have integrated equations (\ref{eqmoto}) on the computer by means of 
a fourth order symplectic algorithm\cite{rf:yo}, with a 
time step fixed in order to have a good energy conservation 
(the average error is of the order ${\Delta E\over E}= 10^{-5}$,
but at very low energy it is necessary to have a higher accuracy and
simulations were done with   ${\Delta E\over E}= 10^{-12}$). 
For each dynamical trajectory we start the system with a given 
initial distribution and we compute ${\theta_i, p_i}$ at each time 
step, and from them the total magnetization $M$ and the temperature 
$T$ (through the relation $T=2<K>/N$). 
We consider systems with various sizes $N$ and we explore a wide 
range of energies $U=E/N$.

\begin{figure}
\epsfxsize = 8 cm 
\epsfysize = 10 cm
\centerline{\epsfbox{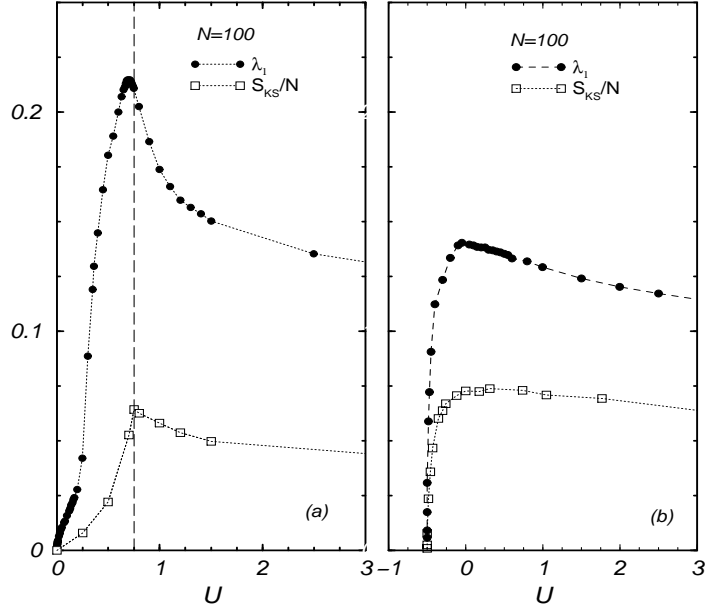}}
\caption{We show the largest Lyapunov exponent $\lambda_1$ 
and the Kolmogorov--Sinai entropy per particle $S_{KS}/N$ as a 
function of $U=E/N$ for the ferromagnetic and the  
antiferromagnetic case. Points are numerical
simulations for N=100, lines are only to guide the eye. The dashed
vertical line indicates the critical point.See text.
}
\label{mari2}
\end{figure}

To characterize chaos we compute the entire Lyapunov spectrum  
$\lambda_i$, $i=1,2,\dots ,2N$ by using the standard method of 
Benettin et al.~\cite{rf:ben,rf:sn}. 
We focus our attention on the Largest Lyapunov Exponent (LLE) 
$\lambda_1$ and on the Kolmogorov--Sinai entropy 
$S_{KS}=\sum^N_{i=1} \lambda_i$, computed as the sum of 
the positive Lyapunov exponents\cite{rf:pesin}    
In fig.2 we report $\lambda_1$ and $S_{KS}/N$ vs. $U$, both for the 
ferromagnetic and the antiferromagnetic case (N=100). 
The HMF is integrable in the two limits of very low and 
very high energy, and this is indifferently true 
for $\epsilon=\pm 1$. 
The LLE and the Kolmogorov--Sinai entropy tend to zero in these two 
limits. 
 
In particular the HMF for $\epsilon=1$ has already been studied in detail 
and the following results have been reported in the literature: 
 
\noindent    
- for $U \to 0$, $\lambda_1 \to 0$ as $U^{\beta}$ where the 
exponent is found to be $\beta=1/2$.
Essentially no dependence on the system size is observed 
in this regime ~\cite{rf:prl1,rf:pd}. 
 
\noindent
- for $U >>  U_c$, $\lambda_1 \to 0$ as $N^{-{1\over 3}}$ and this 
behavior can be explained by means of a random matrix 
approximation\cite{rf:prl1,rf:pari}.  
 
\noindent
The main difference between the ferromagnetic and the 
antiferromagnetic hamiltonian appears at intermediate energy. 
In fact, although both cases are chaotic and we have 
$\lambda_1$ and $S_{KS}/N$ different from zero,  
in the ferromagnetic system one observes a well defined peak 
just below the critical energy $U_c$. In some sense, the 
dynamics feels the presence of the phase transition.   
In fact in ref.\cite{rf:prl1} the increase of the LLE for the 
ferromagnetic case has been related to the increase of kinetic energy 
fluctuations and the specific heat.
It has also been shown that the peak, persists as $N\to \infty $ \cite{rf:pd}. 
This result is also confirmed by a recent more sophisticated theoretical 
calculation~\cite{rf:firpo}, using the formalism introduced in 
ref. ~\cite{rf:pet} .

\begin{figure}
\epsfxsize = 8 cm 
\epsfysize = 10 cm 
\centerline{\epsfbox{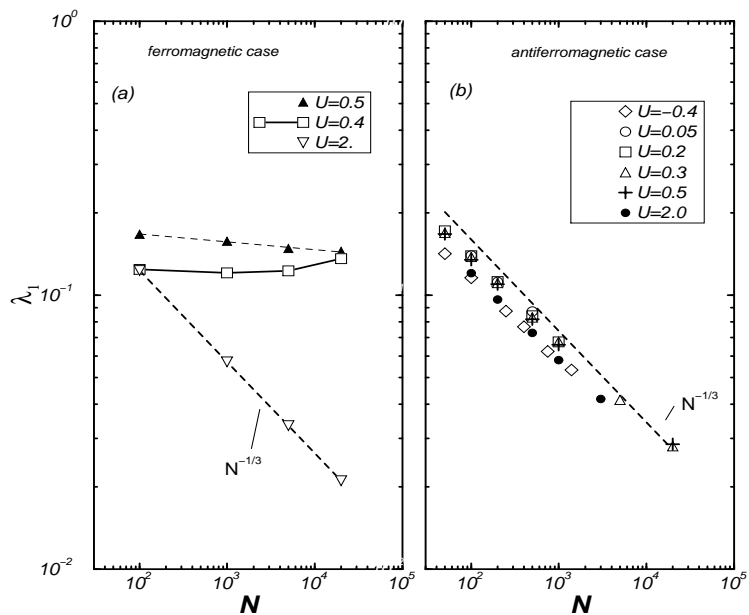}}
\caption{We show the behavior of $\lambda_1$  vs. $N$ for different energies
in the ferromagnetic (a) and in the antiferromagnetic case (b). See text for more
details}
\label{mari3}
\end{figure}

On the contrary, in the antiferromagnetic case a smoothed shoulder 
is found (instead of a peak) both for $\lambda_1$ and for $S_{KS}/N$. 
The difference between the two cases is better visible for $S_{KS}/N$.
In a pioneering paper a similar pronounced peak in LLE 
was found for second--order phase transitions in nuclear-like systems 
~\cite{rf:blr}.
A smooth behavior similar to the one in 
fig.2(b) was found in other models, when there is no phase transition 
in the canonical ensemble~\cite{rf:pet,rf:butera}. 

This different behavior can be better shown by studying 
the LLE as a function of the size $N$ of the system. 
This is displayed in figure 3, where we report $\lambda_1$
as a function of N for several energies below and above the peak. 
In the ferromagnetic case $\lambda_1$ is constant or even increases for 
$U<U_c$, while a decay to zero as $N^{-1/3}$ is evident for $U>U_c$. 
In fact, for  $U>U_c$  the rotators move independently and the 
power law $\lambda_1 \propto N^{-1/3}$ can be well explained by 
a suitable random matrix approximation~\cite{rf:prl1,rf:pari}. 
On the contrary for the antiferromagnetic case fig.3(b), the 
LLE appears to vanish as $N \to \infty$ for all values of $U$
(apart from the very low energy region, where the bi-cluster forms). 
In particular, we find the same  power law $\lambda_1\propto N^{-1/3}$ as 
for the ferromagnetic case in the overcritical region. In fact, 
the random matrix approximation applies also when the potential is repulsive.  
The chaotic behavior is only an artifact of the finite-size fluctuations 
which disappear for $N \rightarrow \infty$. No chaos exist in 
the thermodynamic limit. This behavior strongly contrasts with
what happens for FPU lattices (short-range interactions), where
the LLE reaches a finite value, and therefore chaos persists,
in the thermodynamic limit.

In the ferromagnetic case, close and below the critical point, kinetic 
energy fluctuations are physical and are due to the second--order phase 
transition. The LLE is related to these fluctuations and 
does not go to zero in the thermodynamic limit.

\begin{figure}
\epsfxsize = 8  cm 
\epsfysize = 10 cm 
\centerline{\epsfbox{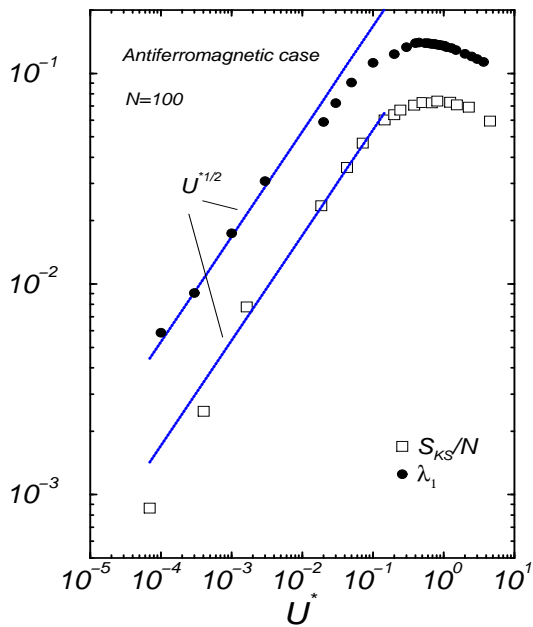}}
\caption{We show
$\lambda_1$ and $S_{KS}/N$ as a function of $U^*=U+0.5$ 
for the antiferromagnetic case in log--log scale. 
Points are microcanonical numerical simulations for N=100. 
Dashed lines are fits of the behavior for very small energies. See text for more
details.} 
\label{mari4}
\end{figure}

In fig. 4 we analyze the behavior of the LLE in the limit of 
small energies for the antiferromagnetic case. 
As we have said previously the equations of the HMF are formally 
equivalent to those of a perturbed pendulum and 
in general we expect the HMF to be integrable when $U$ tends to the 
ground state energy (i.e. when the perturbation goes to zero). 
This has already been checked for the ferromagnetic case~\cite{rf:prl1,rf:pd}
and the law  $\lambda_1\propto U^{1/2} ~,~ U\rightarrow 0$ was found; here we present the 
antiferromagnetic case. 
We show in fig.4 $\lambda_1$ and $S_{KS}/N$ vs. $U^*=U+ 1/2$ 
in log--log scale (the ground state energy in the repulsive case  
is $-1/2$). The dashed line indicates the presence of a power law 
with the same exponent $1/2$ , i.e. $\lambda_1 \sim (U^*)^{1/2} ~,~ U\rightarrow 0$. 
The same exponent was found also for nuclear-like systems \cite{rf:blr}.
Thus, also in this case there seems to be a universal law. However, though
some heuristic arguments have been presented \cite{rf:prl1,rf:pd}, the deep
theoretical reason of this law is not clear. 
 
\section{Conclusions}

We have discussed the most important results recently obtained 
for the HMF model.
This model has revealed very useful for studying chaos
in a Hamiltonian system with many degrees of freedom and in 
particular for undestanding the connection between microscopic
chaos and macroscopic laws, i.e. thermodynamics. We have
also discussed in particular the behavior of the Lyapunov exponents
and the Kolmogorov--Sinai entropy for the ferromagnetic and antiferromagnetic
case. While in the former case, where a second--order phase transition 
is present, one observes a well defined peak in the chaoticity 
indicators ($\lambda_1$ and $S_{KS}/N$) at the critical point, 
in the latter case one has a smoother increase between the two 
integrable limits of very small and very large energy.
In the high energy phase of both the ferromagnetic and the antiferromagnetic
model $\lambda_1$ vanishes as $N^{-1/3}$.
In the ferromagnetic case the peak in $\lambda_1$ persists for 
$N \rightarrow \infty$ and the LLE remains finite in  the whole
low temperature phase. Chaos persist in the thermodynamic limit.
 
Though the border from low-dimensional to more realistic dynamical 
systems has been crossed and a lot of work has been done in this direction
with the help of more powerful computers, we  still  have a long way to go 
in order to understand some important issues at the foundations of 
Statistical Mechanics.
One has to admit that chaos in systems with many degrees
of freedom is still poorly understood and represents the real
challenge for the next decade.

\section*{Acknowledgements}

A.R.  would like to thank the Italian Foreign Office and in particular 
Dr. Sasso for the economic support and Prof. M. Robnik for the invitation 
to participate to this very interesting school/conference and the 
extremely warm hospitality in Maribor. 
 

\end{document}